# Filter-free high-performance single photon emission from a quantum dot in a Fabry-Perot microcavity

Zhixuan Rao[1,†], Jiawei Yang[1,†], Changkun Song[1], Mujie Rao[1], Ziyang Zheng[1], Luyu Liu[1], Xuebin Peng[1], Ying Yu[1,2]*, Siyuan Yu[1,2]

[1]*State Key Laboratory of Optoelectronic Materials and Technologies, School of Electronics and Information Technology, Sun Yat-Sen University, Guangzhou 510006, China*
[2]*Hefei National Laboratory, Hefei 230088, China*
[†]*These authors contributed equally to this work.*
*\*yuying26@mail.sysu.edu.cn*

**Abstract:** Combining resonant excitation with Purcell-enhanced single quantum dots (QDs) stands out as a prominent strategy for realizing high performance solid-state single photon sources. However, optimizing photon efficiency requires addressing challenges associated with effectively separating the excitation laser from QDs' emission. Traditionally, this involves polarization filtering, which limits the achievable polarization directions and the scalability of photonic states. In this study, we have successfully tackled this challenge by employing spatially-orthogonal resonant excitation of QDs, deterministically coupled to monolithic Fabry-Perot microcavities. Leveraging the membrane cavity structures, we have achieved filter-free single photon resonant fluorescence. The resulting source produces single photons with a simultaneous high extraction efficiency of 0.87, purity of 0.9045(4), and indistinguishability of 0.963(4).

## 1. Introduction

Self-assembled quantum dots (QDs) play a crucial role in generating quantum states of light, including single photons [1-3], entangled photon pairs [4-6] and cluster states [7-9]. To fully harness their capabilities, it is essential to controllably prepare photons into a given quantum state with high purity and indistinguishability, as well as extract them efficiently from the semiconductor bulk for collection by a free-space lens or optical fiber. The initial hurdle of efficient preparation has been effectively addressed by employing fast pulsed resonant excitation, eliminating dephasing and time jitter and resulting in near-unity indistinguishable single photons [10]. The subsequent challenge in efficient extraction has been tackled by embedding QDs into Purcell-enhanced photonic cavities, such as micropillar [11-13], photonic waveguide cavity [14,15], and circular Bragg grating [5,16]. The combination of these two strategies produces pure and indistinguishable photons with high extraction efficiency [12-13,15].

However, a critical obstacle to achieving high end-to-end efficiency lies in preventing the excitation laser pulse from entering the collection path. Typically, separating the pump laser from the signal involves pump-probe cross-polarization, leading to a signal reduction of at best 50% when using isotropic cavities [12-13]. Recent advancements have alleviated this issue by developing cavities with degenerate mode-splitting induced by geometry ellipticity [17] or material birefringence [18,19]. In this approach, a circularly polarized quantum emitter is brought into resonance with one of the polarized cavity modes for efficient polarized photon extraction, while the perpendicular polarization is used for laser exclusion. The state-of-the-art QD-based single photon source in open microcavity has achieved high end-to-end efficiency (>71.2%) and high photon indistinguishability (>98.5%) [18,19]. However, optimizing the pulse width remains crucial to balance excitation efficiency and the risk of QD re-excitation [19-21]. Meanwhile, polarized cavities face limitations in generating photons in a fixed polarization, hindering the scaling up of photonic states such as cluster states [6-8].

To overcome this limitation, a straightforward approach involves resonantly exciting the emitter via non-cavity modes with a propagation direction perpendicular to the collection axis, which is commonly used for generating highly indistinguishable photons from planar cavities [22,23] and waveguide-type cavities [15,24]. Despite its attractiveness, implementing this method in high Purcell-enhanced vertical cavities remains challenging due to the isolated structure of micropillars and circular Bragg gratings. Tobias *et al.* attempted to integrate connected waveguides with micropillars, however, it resulted in mode leaking and laser scattering, achieving only a Purcell factor of 2.5 [25]. For open cavities, introducing excitation at another axis on top of the existing two 3D independent adjustments of mirrors is challenging. Here, we propose a feasible solution by introducing a novel microcavity structure that combines a micrometer parabolic lensed-defect between two distributed Bragg reflectors (DBRs). Thanks to the membrane structures, unwanted pump laser scattering is well suppressed by in-plane guided waveguide mode, resulting in a filter-free Rabi oscillation as well as a best autocorrelation value of $g^{(2)}(0)=0.0322(2)$. Furthermore, combining resonant excitation with the Purcell effect enables the generation of single photons with a simultaneous high extraction efficiency of 0.87, purity of 0.9045(4), and indistinguishability of 0.963(4).

## 2. Results

### 2.1 Design concept.

To achieve resonant excitation and effectively discriminate the QD signal from stray laser light, we implemented an orthogonal excitation/detection geometry (Fig. 1(a)). A single mode fiber, mounted on a three-axis displacement stage, is brought within a few microns of the cleaved sample edge. An in-plane polarized tunable mode-locked Ti:Sa pulsed laser is introduced through the fiber to excite the QDs. The emitted QD light, detected without any filters, is coupled into a single-mode fiber. The QD is embedded into a 2-λ monolithic microcavity, consisting of a parabolic lensed-defect within a Fabry-Perot distributed Bragg reflectors (DBR) cavity. Vertical confinement is provided by 7-pair $SiO_2/TiO_2$ top DBRs and 46-pair $GaAs/Al_{0.95}Ga_{0.05}As$ bottom DBRs, while lateral confinement arises from the parabolic lensed-defect in the central spacer layer, as depicted in the cross-section scanning electron microscope (SEM) in Fig. 1(b), and the electric filed distribution of existing planar waveguide mode is inserted. The spatial overlap of the QD emission with the cavity mode is achieved through well-established deterministic positioning techniques [11,26]. Comprehensive fabrication processes are detailed in Supplementary material section 1 and our previous report [27]. Theoretically, an extraction efficiency $η_e$ as high as 94.9% with a Purcell factor of 40 can be attained [27].

In experiments, the laser, directed onto the cleaved side facet of the sample, effectively engages with the planar waveguide mode, enabling low-loss transmission (Figs. 1(c-d)). Notably, our methodology enables QD excitation at a distance greater than 400 μm from the cleaved edge of the sample. The orthogonal alignment between the propagation direction of the resonant excitation laser and the direction of QD emissions minimizes upward scattering, obviating the necessity for polarization filtering. Importantly, our approach offers more versatility than the cross-polarization excitation: the QD emission can exhibit any polarization as desired for specific applications such as BB84 [28] or polarized for other applications such as boson sampling [29].

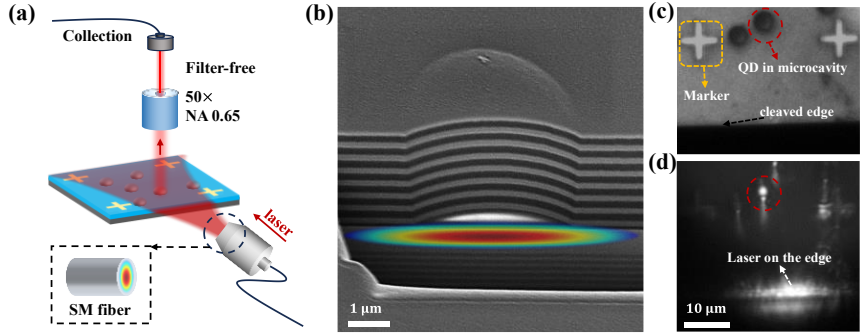

**Fig. 1.** Theoretical framework for spatially-orthogonal excitation. (a) The experimental setup for spatially-orthogonal excitation, involving the precise arrangement of a single-mode optical fiber and the sample within a low-temperature thermostat. The optical fiber, manipulated by a displacement stage with customized mold and metal tape, is positioned a few micrometers from the impeccably cut surface of the sample. Signal light, collected by a coupling lens with a numerical aperture (NA) of 0.65, is oriented perpendicular to the incident laser direction and guided into a single-mode fiber. (b) A scanning electron microscope (SEM) image of the sample, consisting of 7 pairs of $SiO_2/TiO_2$ top DBRs and 46 pairs of $GaAs/Al_{0.95}Ga_{0.05}As$ bottom DBRs. Quantum dots (QDs) are positioned within a parabolic microcavity at the center of the fundamental mode (see Inset). (c) A frontal EMCCD imaging of the sample illustrates the cleaved edge, QD in microcavity, and the marker. (d) A frontal EMCCD imaging captures the spatially-orthogonal excitation of QDs. The incident laser, with specific polarization, couples into the sample, initiating a long-distance, low-loss propagation through waveguide modes and exciting QDs along the propagation path. Noteworthy, markers and other defects within the sample may generate partially negligible scattered light.

## 2.2 Purcell enhanced single photon emission under spatially-orthogonal excitation.

To illustrate our approach, we employ spatially-orthogonal resonant excitation for a single-QD charged exciton (CX) that is deterministically coupled with both circularized and polarized microcavities in two distinct samples. Figs. 2(a-b) show the photoluminescence (PL) of our microcavity modes under above band excitation at high-power, which splits into horizontally- and vertically-polarized (H- and V-polarized) modes, with a separation $\Delta\omega$ of 11.05 GHz in Cavity A and 86.50 GHz in Cavity B. The observed mode splitting is primarily induced by residual asymmetric uniaxial strain in the semiconductor materials, a phenomenon also noted in open cavities [19]. Further adjusting the magnitude of mode-splitting can be achieved by applying strains in the GaAs material [30]. The linewidths of the H mode in Cavity A and Cavity B are 0.098 nm and 0.111 nm, corresponding to measured quality factors of 9511 and 8234, respectively. In birefringent cavity, the spontaneous radiation rate of the exciton's circularly polarized transition is expected to be redistributed into H and V polarizations [18]. By bringing the QD into resonance with the cavity H mode through slight temperature adjustments, predicted polarized spontaneous emission $\zeta_H$ of 0.578 in Cavity A and 0.910 in Cavity B are obtained (see inset of Figs. 2(a-b)).

In the pursuit of creating a highly-bright and indistinguishable single-photon source, we conducted pulsed resonance fluorescence for the CX under laser excitation with an 80.1 MHz repetition rate. Time-resolved resonance fluorescence measurements reveal shortened radiative lifetimes for the QDs on resonance with cavities to $\tau_{on}\sim 134\ ps$ (Cavity A, blue line in Fig. 2(c)) and $\sim 53\ ps$ (Cavity B, red line in Fig. 2(c)), respectively. Compared with the average lifetime for the QDs in bulk from the same area (black line in Fig. 2(c)), a Purcell factor $F_p$ of ~7.5 (19) is achieved in Cavity A (B).

In Fig. 2(d), the detected single photon flux without any filter is presented as a function of the driving laser power, exposing a full Rabi oscillation curve in both types of microcavities, which can be attributed to the coherent control of the two-level system. Notably, this marks the first observation of pulsed resonance fluorescence for QD-in-microcavity systems under spatially-orthogonal excitation. Under the condition referred to as the 'π pulse', an avalanche photon detector (APD) records approximately 4.93 MHz in Cavity A and 6.23 MHz in Cavity B. Accounting the set-up efficiency and the correction factor for the APD, an extraction efficiency of 0.63 is obtained for Cavity A and 0.87 for Cavity B. A comprehensive calibration of the system efficiency is outlined in Supplementary material Section 3. For both circularized and polarized microcavities, spatially-orthogonal excitation proves to be an excellent scheme for resonant fluorescence experiments.

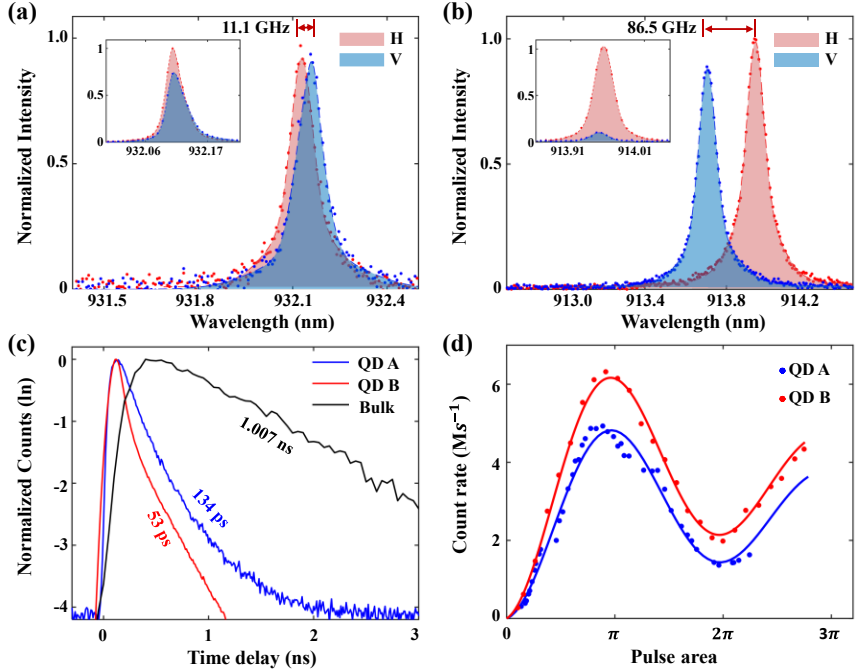

**Fig. 2.** Spatially-orthogonal resonant single-photon characterization of a QD in microcavity. (a) Photoluminescence (PL) spectra of cavity A with a small splitting $\Delta\omega$ =11.05GHz. The H-polarized mode (highlighted in red) is at 932.125 nm with a linewidth of 0.098 nm ($\delta\omega_H$=33.84 GHz), while the V-polarized mode (highlighted in blue) is at 932.157 nm with a linewidth of 0.102 nm ($\delta\omega_V$=35.22 GHz). The inset shows that the H-polarized contribution of the QD is 57.80% while the V-polarized mode is 42.2%. (b) Cavity B exhibits a larger split of $\Delta\omega$ =86.50 GHz. The H-polarized mode is at 913.945 nm with a linewidth of 0.111nm ($\delta\omega_H$=39.84 GHz), and the V-polarized mode is at 913.704 nm with a linewidth of 0.108 nm ($\delta\omega_H$=38.78 GHz). The inset shows the high-polarized single photon, with 91% of QD's emission in the H-mode. (c) Lifetime of resonant fluorescence of QD in microcavity under spatially-orthogonal excitation. Under the 'π pulse' condition, we extracted an ultra-short lifetime from the typical lifetime histogram of CX, which is ~134 ps (blue curve) for QD A in Cavity A and ~53 ps (red curve) for QD B in Cavity B. With the lifetime of ~1.007 ns (black curve) in the bulk material, we obtain the Purcell factor of ~7.5 in cavity A and ~19 in cavity B. (d) The relationship between the pump strength and the count rate of the avalanche photon detector (APD), revealing a full Rabi oscillation curve. The curves represent the numerical fitting results, with the maximum count rate of QD A and QD B being ~4.93 M/s and ~6.23 M/s respectively. The 'π pulse' of QD in cavity A and cavity B corresponds to a power of 66 μW and 36 μW, respectively.

## 2.3 Comparison of spatially-/polarized-orthogonal excitation.

We then assess the single-photon characteristics in Cavity B under both spatially-orthogonal and polarized-orthogonal excitation strategies (Figs. 3(a-b)). For polarized-orthogonal excitation, the dot is excited with V-polarized laser and H polarized single photons are collected. As shown in Fig. 3(b), only about 4.64 million photons per second are detected at π pulse. We estimate the probability of emission into the H-polarized mode, denoted as $\beta_H$, can be determined to be $\beta_H = F_p/(F_p + 1) \times \zeta_H$, revealing a value of 0.865 in Cavity B.

To characterize the purity of the single-photon source, a Hanbury Brown and Twiss setup is employed at the condition of I=0.6·$I_{Sat}$ by decreasing the excitation power, where $I_{Sat}$ represents the intensity of resonant fluorescence at the 'π pulse'. The second-order autocorrelation in Fig. 3(c) reveals a filter-free $g^{(2)}(0)$ of 0.0472(2) under spatially-orthogonal excitation and 0.0889(7) under polarized-orthogonal excitation, indicating both distinct photon antibunching. Furthermore, the coherence of the single photons is assessed using a Hong-Ou-Mandel (HOM) interferometer, with a time separation of 12.48 ns between the two consecutively emitted single photons, consistent with the laser pulses. In Fig. 3(d), the photon correlation histograms of normalized two-photon counts for orthogonal and parallel polarizations indicate a raw HOM visibility of 0.845(2) under spatially-orthogonal excitation and 0.731(4) under polarized-orthogonal excitation. Considering imperfect single-photon purity and an unbalanced beam splitting ratio in the optical setup, a corrected photon indistinguishability of 0.966(2) under spatially-orthogonal excitation and 0.899(4) under polarized-orthogonal excitation is calculated (refer to Supplementary material Section 2). This emphasizes that the single photons generated from QD in the microcavity using spatially-orthogonal excitation exhibit high purity and coherence.

We investigate the origin of the non-vanishing peak at zero-time delay in the QD-in-microcavity system under spatially-orthogonal excitation. In Fig. 3(e), the values of $g^{(2)}(0)$ and HOM visibility ($V_{HOM}$) are presented as functions of the sample temperature. Tuning the QD away from the cavity resonance leads to increased (decreased) lifetime ($F_P$ factor and $V_{HOM}$), indicating that microcavities with high Purcell factors are instrumental in mitigating dephasings. However, the value of $g^{(2)}(0)$ remains at ~0.048, attributed to a slight amount of laser light scattering into the detection channel. This is further supported by the power-dependent behavior of $g^{(2)}(0)$ and HOM visibility in Fig. 3(f), where $g^{(2)}(0)$ increases with higher laser power while the HOM visibility remains almost constant. The best autocorrelation value of $g^{(2)}(0)$ is 0.0322(2) at I=0.2·$I_{Sat}$. The resulting single-photon source exhibits a simultaneous a high extraction efficiency of 0.87, purity of 0.9045(4), and indistinguishability of 0.963(4) at the condition of 'π pulse'. To further alleviate residual laser scattering under spatially-orthogonal excitation, a prospective solution could involve introducing a thicker cavity and surface gold coating.

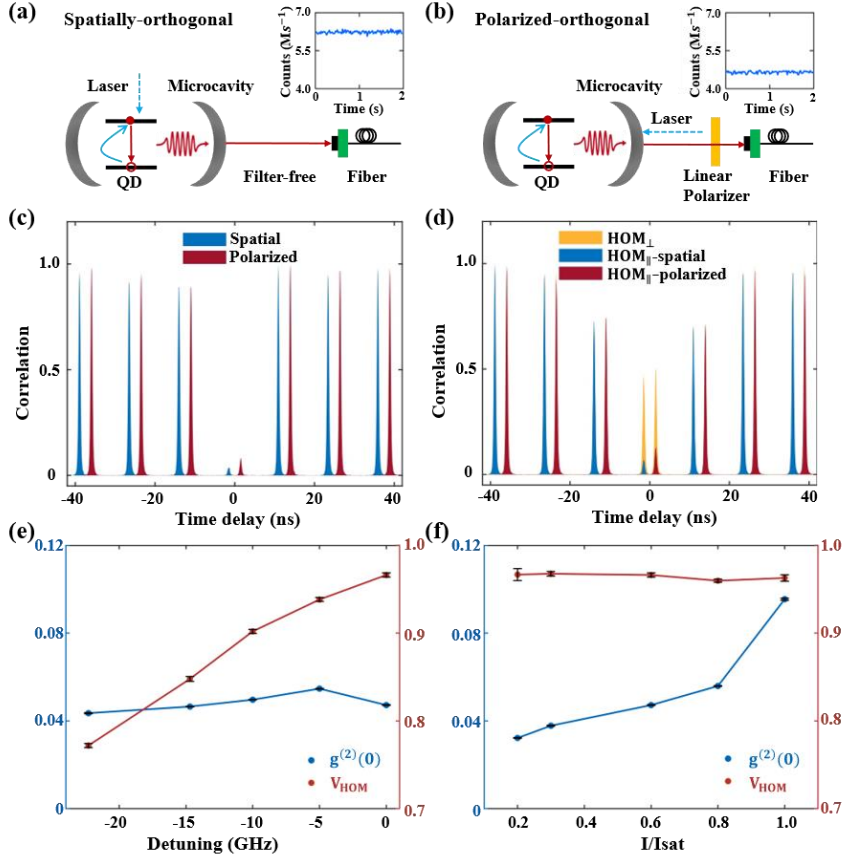

**Fig. 3.** High-performance single photon in microcavity with different excitation strategies. (a) Schematic diagram of spatially-orthogonal excitation. By leveraging the advantage of excitation light propagating orthogonal to the signal direction, filter-free excitation is achieved. The insets depict the saturated count rate of resonant fluorescence of QD CX with a value of 6.23M/s. (b) Schematic diagram of polarized-orthogonal excitation. Orthogonal polarization filtering simultaneously removes a portion of the signal, resulting in the reduction of the saturation count rate to 4.64M/s. (c) Measurement of QD single-photon purity. With the filter-free excitation, we achieved a high-purity single-photon with a value of $g^{(2)}(0)$ =0.0472(2), while it was 0.0889(7) for polarized-orthogonal excitation. (d). Measurement of HOM interference. A clear contrast at zero-time delay of two orthogonal polarization is observed, with values of $g^{(2)}_{\parallel}(0)$ =0.0822(2) and $g^{(2)}_{\perp}(0)$ =0.5312(12), resulting in an indistinguishability of $V_{raw}$=0.845(2) for spatially-orthogonal excitation. For polarized-orthogonal excitation, we attained values of $g^{(2)}_{\parallel}(0)$ =0.1518(3) and $g^{(2)}_{\perp}(0)$ =0.5641(20), indicating $V_{raw}$=0.731(4). (e) Temperature dependence of the purity (blue) and indistinguishability (red). As the temperature increases, the detuning between the QD and the cavity gradually increases, while purity remains stable around 0.952, and indistinguishability decreases gradually. (f) Power dependence of purity (blue) and indistinguishability (red). Purity increases with decreasing power, reaching a maximum of 0.9678(2), influenced by the scattered light while indistinguishability remains stable.

## 2.4 Mollow triplet in the QD-in-microcavity system under continuous-wave spatially-orthogonal excitation.

Due to the membrane structures, the laser utilized for resonant excitation effectively couples with in-plane guided waveguide mode while concurrently suppressing background scattering. Consequently, we also transition to continuous-wave resonance excitation to observe the Mollow triplet, a distinctive feature indicating resonance fluorescence from a strongly driven and dressed two-level system. Fig. 4(a) illustrates a comprehensive power-dependent QD PL series, revealing the emergence of two symmetrically spaced sidebands around the central line as power increases. The extracted Rabi splitting is graphed against the square root of laser power in Fig. 4(b), validating the theoretically expected linear proportionality, with a significant Rabi splitting of up to 21.4 GHz. Fig. 4(c) demonstrates the linewidth of the Rabi sidebands, exhibiting a clear linear dependence on the squared Rabi frequency. These linear trends strongly suggest phonon-dressed QD Mollow triplet emission within the cavity-quantum electrodynamics (cQED) regime in our sample [31].

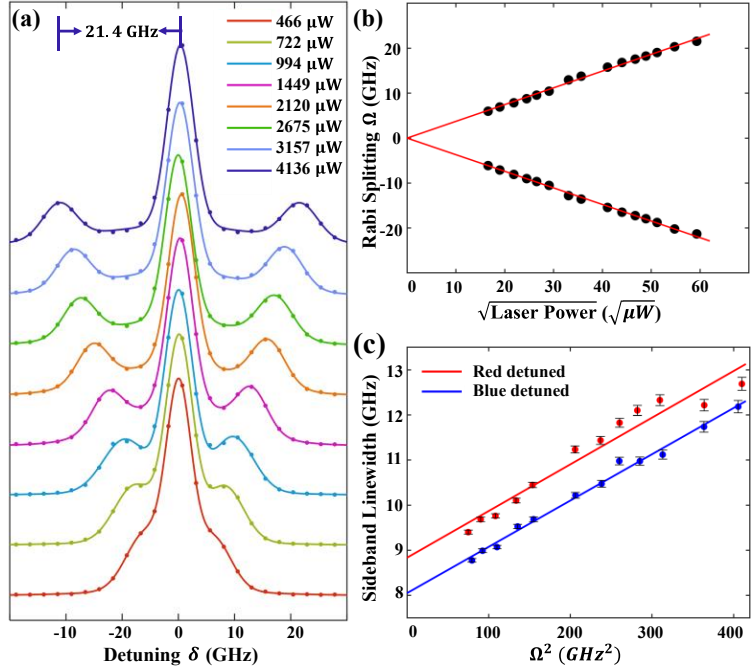

**Fig. 4.** Mollow triplet of QD in microcavity under spatially-orthogonal excitation. (a) A sequence of high-resolution PL spectra, varying with laser power, were obtained using continuous-wave laser resonant excitation of QD CX under spatially-orthogonal excitation. A maximum splitting of 21.4 GHz was achieved at a power of 4.1mW. (b) The correlation between the detuning of the two sideband peaks (Rabi splitting) and the square root of laser power is extracted from panel a, revealing a strictly linear relationship. (c) The association between the linewidths of the two side peaks and the square of the Rabi splitting, revealing linear relationships for both.

## 3. Discussion and conclusion

In summary, our study combines spatially-orthogonal resonant excitation with Purcell-enhanced single QDs in a monolithic Fabry-Perot microcavity, demonstrating the realization of filter-free single photon resonant fluorescence. By addressing the challenges associated with separating the excitation laser from QDs' emission in microcavity, our spatially-orthogonal excitation approach eliminates the requirement for polarization filtering, offering flexibility in controlling the polarization of QD emission. These characteristics collectively establish the high-performance attributes of our single photon source, featuring an efficiency of 87%, a purity of 90.5% and an indistinguishability of 96.3%. Furthermore, we elucidate a strongly driven and dressed two-level system by the observation of a Mollow triplet with a substantial 21.4 GHz Rabi splitting. In future endeavors, our work sets the stage for achieving filter-free single-photon emission with high efficiency and versatility, opening avenues for the development of more efficient and miniaturized quantum light sources. These sources could be readily integrated with fibers by direct attachment to a fiber facet [32] or transferred onto the grating of photonic integrated circuits [33].

**Acknowledgments**

We are grateful for financial support from the Science and Technology Program of Guangzhou (202103030001), the Innovation Program for Quantum Science and Technology (2021ZD0301400), the National Key R&D Program of Guang-dong Province (2020B0303020001), the National Natural Science Foundation of China (12074442).

**Disclosures**

The authors declare no conflicts of interest.

**References**

1. I. Aharonovich, D. Englund, and M. Toth, "Solid-state single-photon emitters," Nat. Photonics 10, 631–641 (2016).
2. P. Senellart, G. Solomon, and A. White, "High-performance semiconductor quantum-dot single-photon sources," Nat. Nanotechnol. 12, 1026–1039 (2017).
3. C. Y. Lu and J. W. Pan, "Quantum-dot single-photon sources for the quantum internet," Nat. Nanotechnol. 16, 1294–1296 (2021).
4. J. Wang, M. Gong, G. C. Guo, and L. He, "Towards Scalable Entangled Photon Sources with Self-Assembled InAs/GaAs Quantum Dots," Phys. Rev. Lett. 115, 67401 (2015).
5. J. Liu, R. Su, Y. Wei, B. Yao, S. F. C. da Silva, Y. Yu, J. Iles-Smith, K. Srinivasan, A. Rastelli, J. Li, and X. Wang, "A solid-state source of strongly entangled photon pairs with high brightness and indistinguishability," Nat. Nanotechnol. 14, 586–593 (2019).
6. K. D. Zeuner, K. D. Jöns, L. Schweickert, C. Reuterskiöld Hedlund, C. Nuñez Lobato, T. Lettner, K. Wang, S. Gyger, E. Schöll, S. Steinhauer, M. Hammar, and V. Zwiller, "On-Demand Generation of Entangled Photon Pairs in the Telecom C-Band with InAs Quantum Dots," ACS Photonics 8, 2337–2344 (2021).
7. I. Schwartz, D. Cogan, E. R. Schmidgall, Y. Don, L. Gantz, O. Kenneth, N. H. Lindner, and D. Gershoni, "Deterministic generation of a cluster state of entangled photons," Science (80-. ). 354, 434–437 (2016).
8. D. Cogan, Z. E. Su, O. Kenneth, and D. Gershoni, "Deterministic generation of indistinguishable photons in a cluster state," Nat. Photonics 17, 324–329 (2023).
9. N. Coste, D. A. Fioretto, N. Belabas, S. C. Wein, P. Hilaire, R. Frantzeskakis, M. Gundin, B. Goes, N. Somaschi, M. Morassi, A. Lemaître, I. Sagnes, A. Harouri, S. E. Economou, A. Auffeves, O. Krebs, L. Lanco, and P. Senellart, "High-rate entanglement between a semiconductor spin and indistinguishable photons," Nat. Photonics 17, 582–587 (2023).
10. Y. M. He, Y. He, Y. J. Wei, D. Wu, M. Atatüre, C. Schneider, S. Höfling, M. Kamp, C. Y. Lu, and J. W. Pan, "On-demand semiconductor single-photon source with near-unity indistinguishability," Nat. Nanotechnol. 8, 213–217 (2013).
11. S. Liu, Y. Wei, R. Su, R. Su, B. Ma, Z. Chen, H. Ni, Z. Niu, Y. Yu, Y. Wei, X. Wang, and S. Yu, "A deterministic quantum dot micropillar single photon source with >65% extraction efficiency based on fluorescence imaging method," Sci. Rep. 7, 1–8 (2017).
12. X. Ding, Y. He, Z.-C. Duan, N. Gregersen, M.-C. Chen, S. Unsleber, S. Maier, C. Schneider, M. Kamp, S. Höfling, C.-Y. Lu, and J.-W. Pan, "On-Demand Single Photons with High Extraction Efficiency and Near-Unity Indistinguishability from a Resonantly Driven Quantum Dot in a Micropillar," Phys. Rev. Lett. 116, 20401 (2016).
13．N. Somaschi, V. Giesz, L. De Santis, J. C. Loredo, M. P. Almeida, G. Hornecker, S. L. Portalupi, T. Grange, C. Antón, J. Demory, C. Gómez, I. Sagnes, N. D. Lanzillotti-Kimura, A. Lemaítre, A. Auffeves, A. G. White, L. Lanco, and P. Senellart, "Near-optimal single-photon sources in the solid state," Nat. Photonics 10, 340–345 (2016).
14. F. Liu, A. J. Brash, J. O'Hara, L. M. P. P. Martins, C. L. Phillips, R. J. Coles, B. Royall, E. Clarke, C. Bentham, N. Prtljaga, I. E. Itskevich, L. R. Wilson, M. S. Skolnick, and A. M. Fox, "High Purcell factor generation of indistinguishable on-chip single photons," Nat. Nanotechnol. 13, 835–840 (2018).
15. R. Uppu, F. T. Pedersen, Y. Wang, C. T. Olesen, C. Papon, X. Zhou, L. Midolo, S. Scholz, A. D. Wieck, A. Ludwig, and P. Lodahl, "Scalable integrated single-photon source," Sci. Adv. 6, 1–6 (2020).
16. M. Moczała-Dusanowska, Ł. Dusanowski, O. Iff, T. Huber, S. Kuhn, T. Czyszanowski, C. Schneider, and S. Höfling, "Strain-Tunable Single-Photon Source Based on a Circular Bragg Grating Cavity with Embedded Quantum Dots," ACS Photonics 7, 3474–3480 (2020).
17. H. Wang, Y. M. He, T. H. Chung, H. Hu, Y. Yu, S. Chen, X. Ding, M. C. Chen, J. Qin, X. Yang, R. Z. Liu, Z. C. Duan, J. P. Li, S. Gerhardt, K. Winkler, J. Jurkat, L. J. Wang, N. Gregersen, Y. H. Huo, Q. Dai, S. Yu, S. Höfling, C. Y. Lu, and J. W. Pan, "Towards optimal single-photon sources from polarized microcavities," Nat. Photonics 13, 770–775 (2019).
18. N. Tomm, A. Javadi, N. O. Antoniadis, D. Najer, M. C. Löbl, A. R. Korsch, R. Schott, S. R. Valentin, A. D. Wieck, A. Ludwig, and R. J. Warburton, "A bright and fast source of coherent single photons," Nat. Nanotechnol. 16, 399–403 (2021).
19. X. Ding, Y.-P. Guo, M.-C. Xu, R.-Z. Liu, G.-Y. Zou, J.-Y. Zhao, Z.-X. Ge, Q.-H. Zhang, H.-L. Liu, M.-C. Chen, H. Wang, Y.-M. He, Y.-H. Huo, C.-Y. Lu, and J.-W. Pan, "High-efficiency single-photon source above the loss-tolerant threshold for efficient linear optical quantum computing," arXiv:2311.08347 (2023).
20. A. Javadi, N. Tomm, N. O Antoniadis, A. J Brash, R. Schott, S. R Valentin, A. D Wieck, A. Ludwig, R. J Warburton, "Cavity-enhanced excitation of a quantum dot in the picosecond regime," New Journal of Physics 25: 093027 09 (2023).
21. L. Hanschke, K. A. Fischer, S. Appel, D. Lukin, J. Wierzbowski, S. Sun, R. Trivedi, J. Vučković, J. J. Finley, and K. Müller, "Quantum dot single-photon sources with ultra-low multi-photon probability," npj Quantum Inf. 4, 43 (2018).
22. A. Muller, E. B. Flagg, P. Bianucci, X. Y. Wang, D. G. Deppe, W. Ma, J. Zhang, G. J. Salamo, M. Xiao, and C. K. Shih, "Resonance fluorescence from a coherently driven semiconductor quantum dot in a cavity," Phys. Rev. Lett. 99, 2–5 (2007).
23. J. Robertson, S. Founta, M. Hughes, M. Hopkinson, A. J. Ramsay, M. S. Skolnick, and C. K. Shih, "Polarization-resolved resonant fluorescence of a single semiconductor quantum dot," Appl. Phys. Lett. 101, 2010–2014 (2012).
24. S. Kalliakos, Y. Brody, A. J. Bennett, D. J. P. Ellis, J. Skiba-Szymanska, I. Farrer, J. P. Griffiths, D. A. Ritchie, and A. J. Shields, "Enhanced indistinguishability of in-plane single photons by resonance fluorescence on an integrated quantum dot," Appl. Phys. Lett. 109, 151112 (2016).
25. T. Huber, M. Davanco, M. Müller, Y. Shuai, O. Gazzano, and G. S. Solomon, "Filter-free single-photon quantum dot resonance fluorescence in an integrated cavity-waveguide device," Optica 7, 380 (2020).
26. S. Liu, K. Srinivasan, and J. Liu, "Nanoscale Positioning Approaches for Integrating Single Solid-State Quantum Emitters with Photonic Nanostructures," Laser Photonics Rev. 15, 1–15 (2021).
27. J. Yang, Y. Chen, Z. Rao, Z. Zheng, C. Song, Y. Chen, K. Xiong, P. Chen, C. Zhang, W. Wu, Y. Yu, and S. Yu, "Tunable quantum dots in monolithic Fabry-Perot microcavities for high-performance single-photon sources," Light: Science & Applications, 13, 33 (2024).
28. C. H. Bennett and G. Brassard, "Quantum cryptography: Public key distribution and coin tossing," Theor. Comput. Sci. 560, 7–11 (2014).
29. S. Aaronson and A. Arkhipov, "The Computational Complexity of Linear Optics," in STOC 11: PROCEEDINGS OF THE 43RD ACM SYMPOSIUM ON THEORY OF COMPUTING, Annual ACM Symposium on Theory of Computing (2011), pp. 333–342.
30. N. Tomm, A. R. Korsch, A. Javadi, D. Najer, R. Schott, S. R. Valentin, A. D. Wieck, A. Ludwig, and R. J. Warburton, "Tuning the Mode Splitting of a Semiconductor Microcavity with Uniaxial Stress," Phys. Rev. Appl. 15, 1 (2021).
31. S. M. Ulrich, S. Ates, S. Reitzenstein, A. Löffler, A. Forchel, and P. Michler, "Dephasing of triplet-sideband optical emission of a resonantly driven InAs/GaAs quantum dot inside a microcavity," Phys. Rev. Lett. 106, 1–4 (2011).
32. H. Snijders, J. A. Frey, J. Norman, V. P. Post, A. C. Gossard, J. E. Bowers, M. P. Van Exter, W. Löffler, and D. Bouwmeester, "Fiber-Coupled Cavity-QED Source of Identical Single Photons," Phys. Rev. Appl. 9, 31002 (2018).
33. J. Goyvaerts, A. Grabowski, J. Gustavsson, S. Kumari, A. Stassen, R. Baets, A. Larsson, and G. Roelkens, "Enabling VCSEL-on-silicon nitride photonic integrated circuits with micro-transfer-printing," Optica 8, 1573 (2021).